# Random-h Fractional-Dimensional Lattices Reveal Endpoint-Compressed Percolation Activation between Two and Three Dimensions

Ran Huang
Centre of Biomedical Engineering and Technology, Yiwu Research Institute of Fudan University, Yiwu 322000, China
*Correspondence: huangran@fudan.edu.cn

## Abstract

Non-integer dimensionality is central to fractal and complex systems, yet it is rarely represented as an explicit lattice on which classical statistical-mechanical models can be directly simulated. Here we introduce random-h fractional dimension (RhFD), a constructive lattice framework in which fractional-dimensional environments are generated by stochastic activation of local connectivity, $h$. In the 2D-to-3D interval, RhFD lattices are formed by recursively growing out-of-plane sites from a square base with probability $\rho_h$. Using quenched site-percolation simulations, we show that the construction recovers the integer-dimensional endpoints and yields a robust crossover in which the percolation threshold decreases from the 2D regime toward the 3D regime. The crossover is not a uniform interpolation: high-resolution scans reveal endpoint-compressed activation, with $-dp_c/d\rho_h$ increasing toward $\rho_h = 1$. Mass dimension increases with $\rho_h$, whereas the coordination descriptor first decreases as sparse protrusions form and then rises sharply when a dense 3D backbone emerges. RhFD provides an explicit lattice substrate for fractional-dimensional statistical mechanics and shows that geometric mass, local coordination, and critical connectivity can decouple during dimensional crossover.

## 1. Introduction

Dimensionality is one of the most fundamental control variables in statistical mechanics. Critical thresholds, universality classes, correlation functions, and scaling exponents depend strongly on the spatial dimension of the underlying lattice or continuum [1-3]. Classical examples include the difference between one-dimensional and higher-dimensional percolation, the special role of two-dimensional lattice models, and the changing critical behavior of spin and connectivity models across dimensions [4-6]. Dimension is therefore not merely a geometric label; it determines how local interactions accumulate into global connectivity and criticality.

Non-integer dimensionality has long appeared in theoretical physics and mathematics. In field-theoretic approaches, dimension can be analytically continued, as in epsilon-expansion methods [7,8]. In geometry and complex systems, fractal dimension quantifies self-similar or scale-dependent structures [9,10]. These ideas are powerful, but they do not automatically define a simple physical lattice continuously intermediate between, for example, a square lattice and a cubic lattice. A fractal object may have non-integer dimension while possessing connectivity and local coordination very different from ordinary lattices [11-13]. Conversely, analytic continuation in dimension does not specify which sites exist, which sites are neighbors, or how a classical lattice process should be simulated on a concrete geometry.

A complementary route is to construct non-integer-dimensional lattices explicitly. Such a construction should recover standard integer-dimensional endpoints, define actual sites and nearest-neighbor connections, support finite-size simulations, and allow geometric and physical descriptors to be measured rather than imposed. Percolation is a natural first test because it directly probes global connectivity without requiring a Hamiltonian or energetic assumptions [4-6]. If a fractional-dimensional lattice is physically meaningful, its percolation behavior should connect known integer-dimensional limits in a systematic and interpretable way.

Here we introduce random-h fractional dimension (RhFD), a lattice construction based on stochastic activation of local connectivity. The symbol $h$ denotes the local coordination environment of a site; unlike a fixed coordination number in a regular lattice, $h$ varies across the quenched random geometry and is used here to emphasize the heterogeneous connectivity generated by random growth. In the present 2D-to-3D realization, an L × L square lattice is used as the base layer, and additional out-of-plane sites are recursively generated with probability $\rho_h$. This produces quenched random slabs with heterogeneous column heights. At $\rho_h = 0$, the system is a pure two-dimensional square lattice; at $\rho_h = 1$, it becomes a fully occupied 3D slab, with the full cubic case used as the three-dimensional baseline. Intermediate $\rho_h$ values define explicit random lattices between these endpoints.

Using quenched site-percolation simulations, we show that RhFD recovers the integer-dimensional endpoints and produces a robust 2D-to-3D connectivity crossover. The percolation threshold decreases systematically with out-of-plane growth, but the decrease is not uniform. Effective three-dimensional percolative behavior is activated in an endpoint-compressed manner near $\rho_h = 1$: geometric mass accumulates first, whereas a connected three-dimensional backbone emerges only when out-of-plane growth is nearly saturated. By comparing mass dimension $D_m$, coordination descriptor $D_h$, and percolation threshold $p_c$, we show that distinct physical meanings of dimension can separate within a random fractional-dimensional lattice.

## 2. Results

### 2.1. Construction of random-h fractional-dimensional lattices

To generate explicit non-integer-dimensional lattices, we constructed random-h fractional-dimensional lattices by stochastic out-of-plane growth from a two-dimensional square base. All base-layer sites ($z = 0$) were present, and sites at layer $z > 0$ existed only if the site directly below was present and a Bernoulli trial with growth probability $\rho_h$ succeeded. This produces quenched lattices with heterogeneous column heights, preserving nearest-neighbor connectivity along $x$, $y$, and $z$. The mass dimension $D_m$, coordination descriptor $D_h$, and percolation threshold $p_c$ were computed for each lattice to quantify geometric and connectivity properties.

Figure 1 shows a four-panel visualization illustrating the evolution from sparse early out-of-plane growth ($\rho_h$ = 0.20) to a forest-like intermediate structure ($\rho_h$ = 0.55), a denser intermediate state ($\rho_h$ = 0.85), and the fully occupied three-dimensional slab ($\rho_h$ = 1.00). Per-panel metrics $D_m$, $D_h$, and mean column height are indicated.

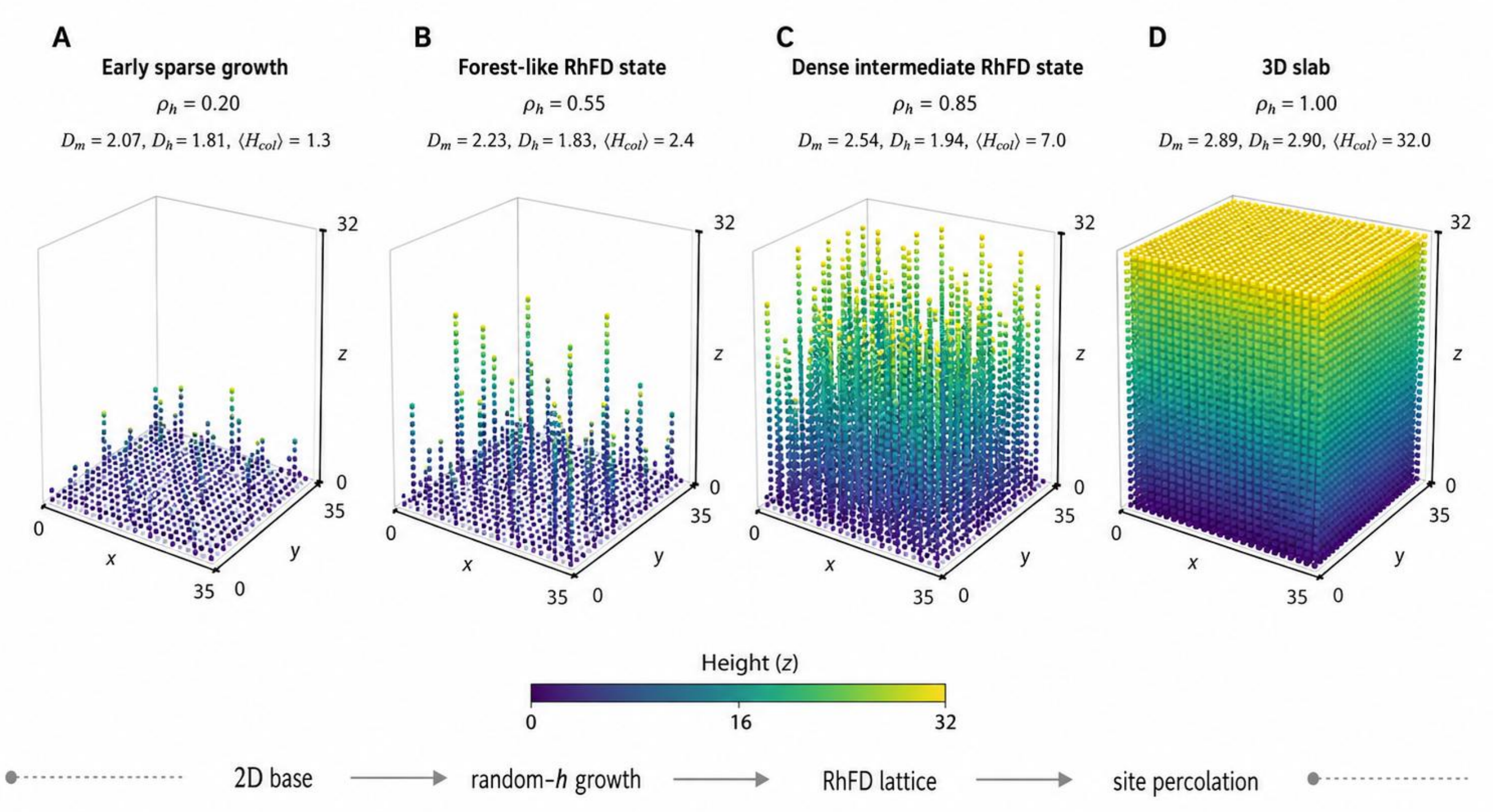

Figure 1. Evolution from sparse out-of-plane growth to a three-dimensional slab. Random-h fractional-dimensional lattices are generated by recursive out-of-plane growth from a two-dimensional base. The four panels show representative geometries at increasing $\rho_h$. The color encodes height z. The metrics $D_m$, $D_h$, and <$H_{col}$> summarize mass dimension, coordination descriptor, and mean column height.

## 2.2. Integer-dimensional endpoint validation

We first validated the framework in the two integer-dimensional limits. In the two-dimensional baseline, the spanning probability $P_{perc}(p)$ sharpened with system size, and the estimated threshold approached the classical square-lattice site-percolation value $p_c^{2D}$ approximately 0.5927 [14,15]. In the three-dimensional simple-cubic baseline, finite-size thresholds decreased from $p_c$ = 0.3242 at L=16 to $p_c$ = 0.3174 at L = 32, approaching the known three-dimensional threshold [15,16]. These endpoints confirm that the percolation detection and threshold extraction correctly reproduce standard lattice limits, establishing a foundation for interpreting RhFD intermediate states.

Figure 2 presents these endpoints. Panels A and B show $P_{perc}(p)$ for two-dimensional and three-dimensional baselines, and Panels C and D display finite-size threshold estimates.

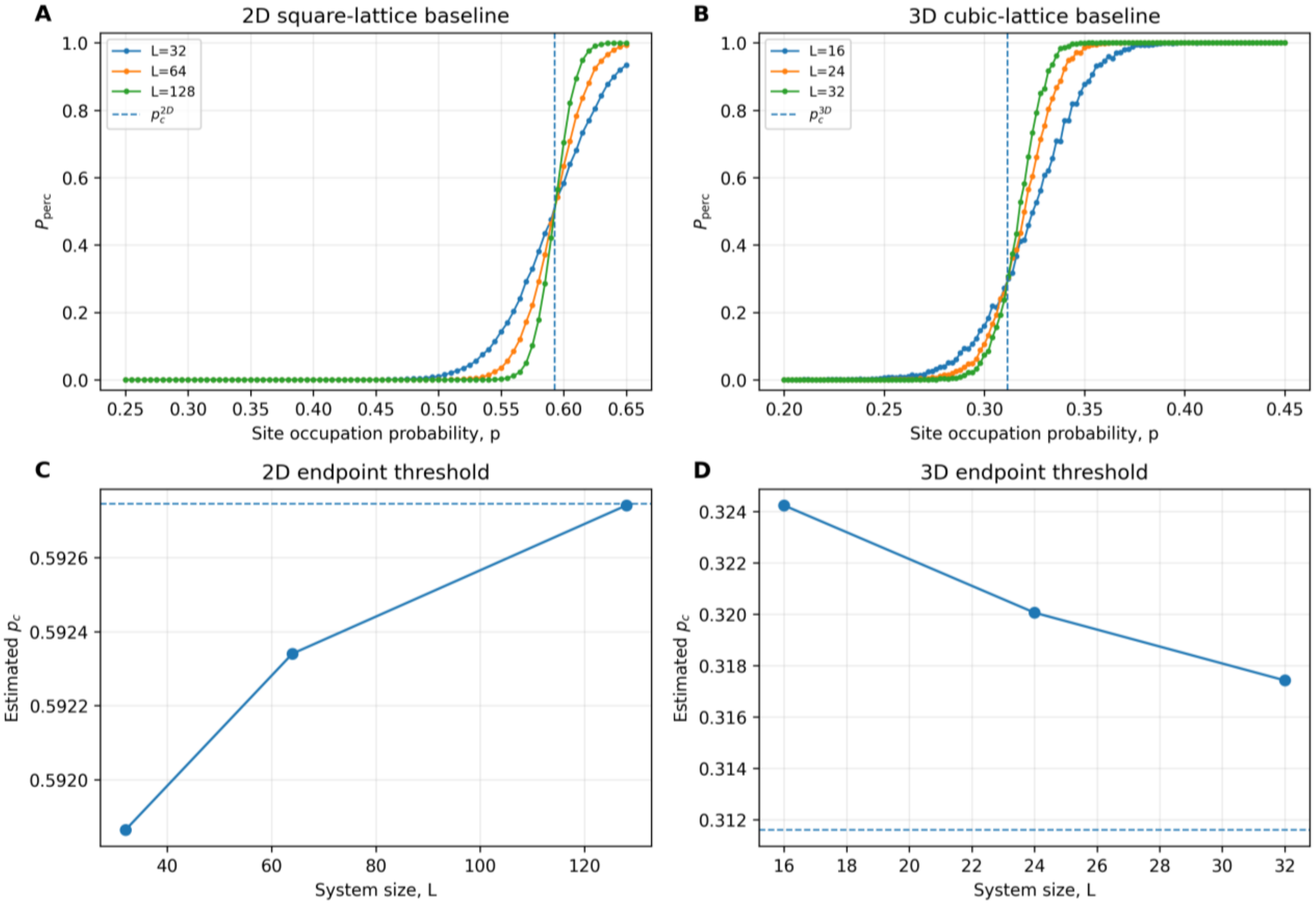

Figure 2. Integer-dimensional validation of the RhFD percolation framework. (A) Site-percolation spanning probability on pure two-dimensional square lattices. (B) Site-percolation spanning probability on pure three-dimensional simple-cubic lattices. (C) Estimated finite-size thresholds for the two-dimensional baselines. (D) Estimated finite-size thresholds for the three-dimensional baselines.

## 2.3. 2D-to-3D crossover in RhFD lattices

RhFD lattices were generated with varying $\rho_h$, and site-percolation simulations were performed using a quenched-disorder protocol: for each fixed lattice geometry, $N_O$ independent occupation realizations were averaged, then averaged over $N_G$ independent geometries. Percolation was defined as the existence of at least one cluster spanning the lattice from $x = 0$ to $x = \text{L-1}$ [14,17].

As $\rho_h$ increased, the percolation threshold $p_c(\rho_h)$ decreased monotonically, reflecting enhanced connectivity from out-of-plane growth. For L = 128, $p_c$ decreased from 0.5927 at $\rho_h = 0$ to 0.3215 at $\rho_h = 1.00$. The close agreement between L = 64 and L = 128 indicates that the observed crossover is not a small-system artifact.

Figure 3 shows the main crossover. Panels A and B display representative $P_{\text{perc}}(p)$ curves for L = 64 and L = 128, Panel C shows $p_c$ ($\rho_h$), and Panel D shows peak positions of the finite-cluster-size response.

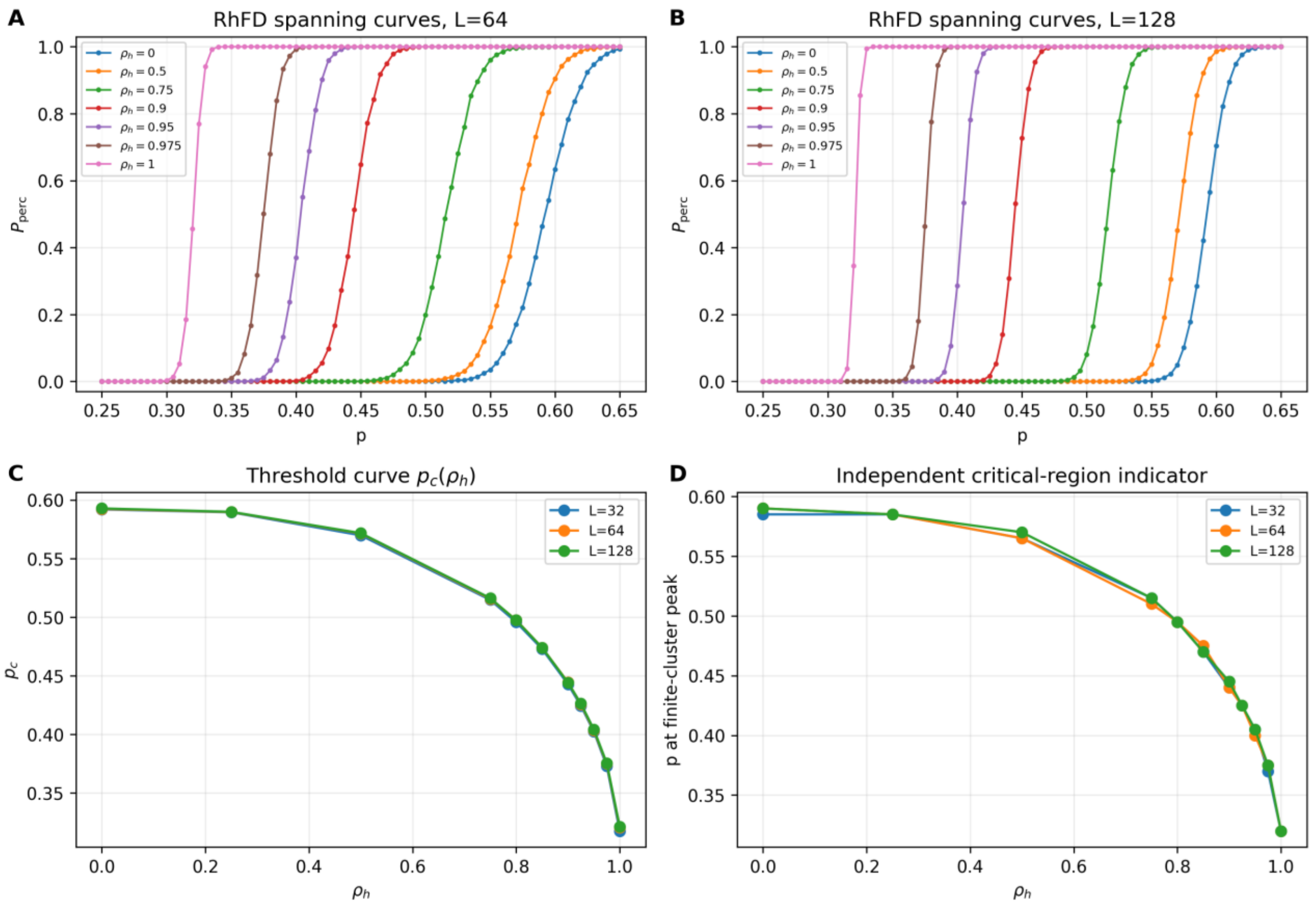

Figure 3. Site-percolation crossover on random-h fractional-dimensional lattices. Increasing $\rho_h$ shifts the spanning-probability curves toward lower occupation probability. The extracted threshold $p_c$ ($\rho_h$) decreases monotonically from the two-dimensional regime toward the three-dimensional slab-like regime.

## 2.4. Endpoint-compressed activation

High-resolution scans in the high-growth regime ($\rho_h$ =0.90-1.00) revealed that $p_c(\rho_h)$ decreases gradually at lower $\rho_h$ but rapidly near the three-dimensional endpoint. The activation susceptibility was defined as

$$\chi_{\text{act}}(\rho_h) = -dp_c/\rho_h.$$

For both L = 64 and L = 128, $\chi_{\text{act}}$ increased toward $\rho_h$ = 1.00. The largest interior response occurred at $\rho_h$ = 0.995, whereas the global maximum occurred at the endpoint. This indicates that effective three-dimensional connectivity emerges primarily when out-of-plane growth is nearly saturated.

Figure 4 illustrates this endpoint-compressed activation. Panel A shows $p_c(\rho_h)$, Panel B shows $\chi_{\text{act}}$, Panel C zooms in on the final interval, and Panel D summarizes global versus interior activation maxima.

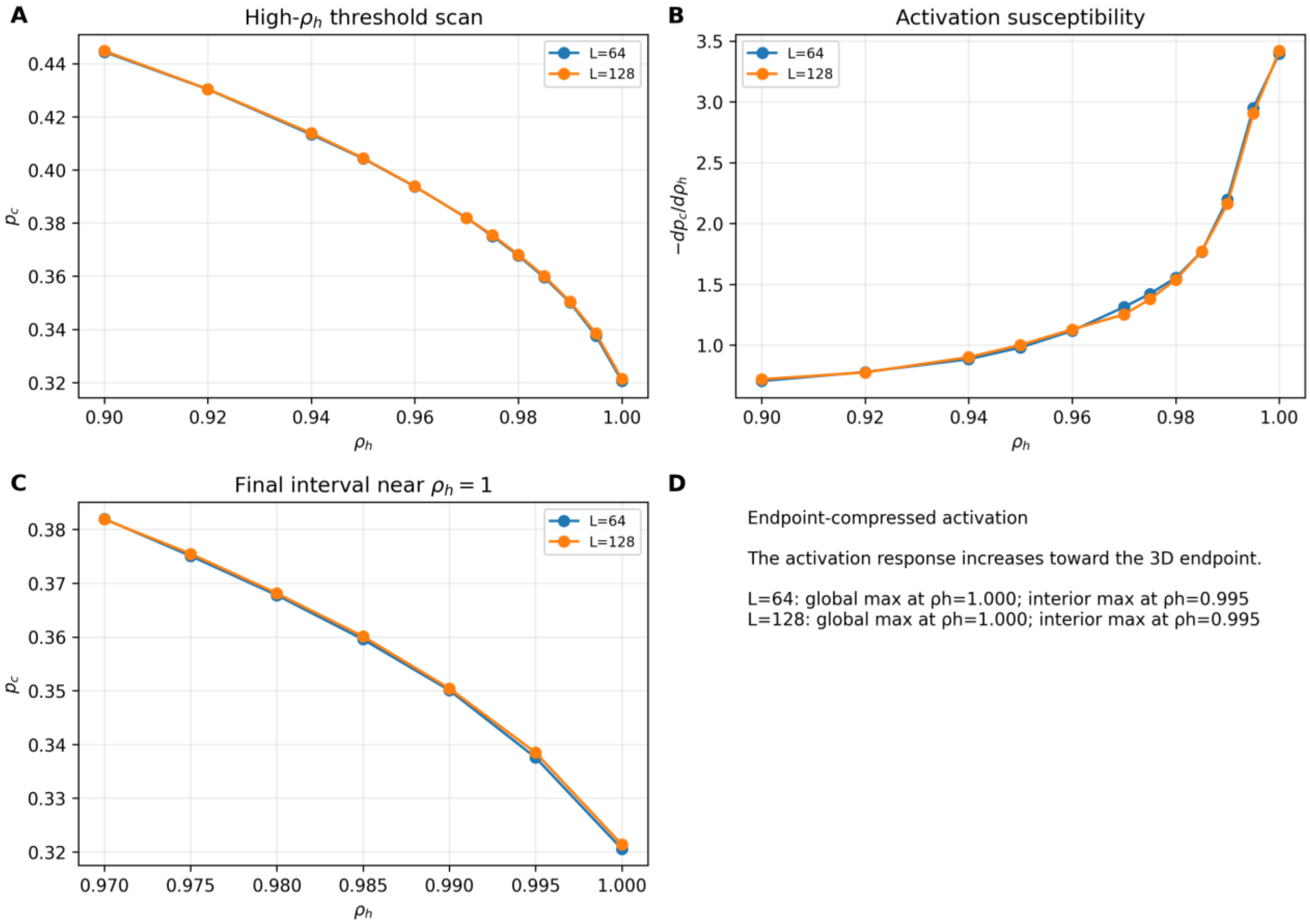

Figure 4. Endpoint-compressed activation of percolative dimensionality. High-resolution scans show that the threshold descent accelerates near $\rho_h$ = 1. The activation susceptibility $-dp_c/d\rho_h$ increases toward the endpoint, indicating non-uniform activation of effective three-dimensional connectivity.

## 2.5. Geometry descriptors and delayed activation

To explain endpoint-compressed activation, we examined mass dimension $D_m = \log N/\log L$, coordination descriptor $D_h = \langle h\rangle/2$, mean column height, and layer occupation profile $\Theta_z = N_z/L^2$. At low to intermediate $\rho_h$, $D_m$ increases gradually while $D_h$ can decrease because sparse protrusions add sites without forming a dense backbone. Only near high $\rho_h$ does $D_h$ rise sharply and layer occupation saturates, producing a connected three-dimensional structure.

Figure 5 summarizes this mechanism. Panels A-C show $D_m$, $D_h$, and mean column height; Panel D shows layer occupation profiles for selected $\rho_h$ values. The profiles show that high layers become substantially occupied only near the three-dimensional endpoint.

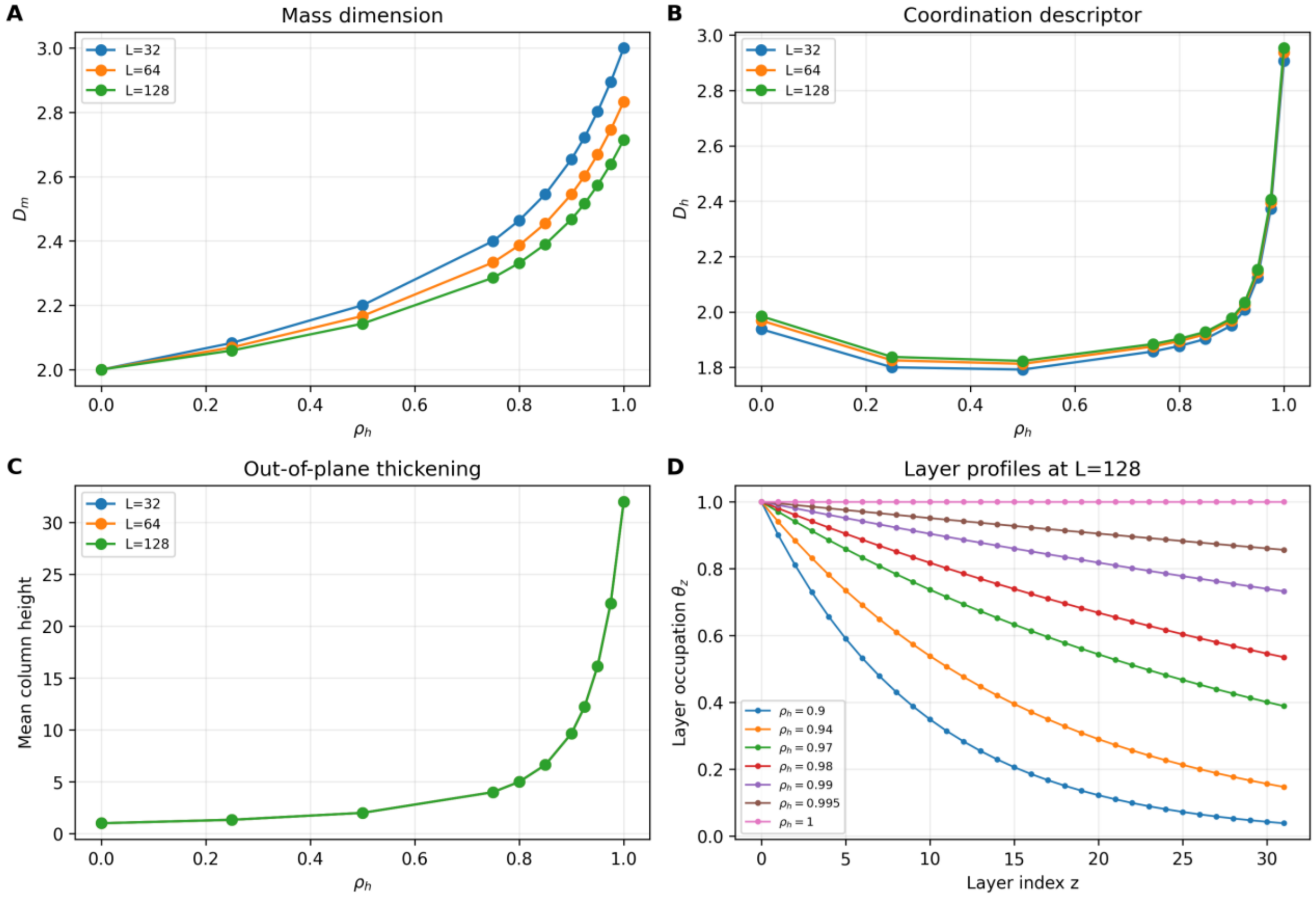

Figure 5. Geometric origin of delayed percolative activation. Mass dimension and mean column height increase with $\rho_h$, while the coordination descriptor first decreases and then rises sharply. Layer occupation profiles show delayed formation of a dense three-dimensional backbone.

# 3. Methods

## 3.1. Random-h fractional-dimensional lattice construction

RhFD lattices are constructed explicitly by stochastic out-of-plane growth from a two-dimensional square base of size L × L at layer $z = 0$. All base-layer sites are present. Sites at layer $z > 0$ exist if the site below exists and a Bernoulli trial with growth probability $\rho_h$ succeeds:

$$s_{x,y,0} = 1, s_{x,y,z} = s_{x,y,z-1}\big(U_{z,y,x} < \rho_h\big), z = 1, \dots, H-1.$$

Here $U[z,y,x]$ is a pre-generated uniform random field in [0,1]. This recursive rule produces quenched lattices with heterogeneous column heights while preserving nearest-neighbor connectivity along $x$, $y$, and $z$. Key structural descriptors were measured for each lattice: mass dimension

$$D_m = \log N / \log L;$$

coordination descriptor

$$D_h = \langle h \rangle / 2,$$

where h is the number of existing nearest neighbors per site; and column-height metrics including mean height, maximum height, and layer occupancy

$$\theta_z = N_z / L^2.$$

### 3.2. Site-percolation simulations

For each generated lattice, site percolation was performed by occupying each site independently with probability $p$. A realization was counted as spanning if at least one connected cluster touched both opposite boundaries in the $x$ direction, from $x = 0$ to $x = L-1$. For two-dimensional lattices, this corresponds to left-to-right boundary spanning; for three-dimensional lattices and RhFD geometries, the cluster may connect any available y position and any existing $z$ layer. This boundary-to-boundary criterion avoids artifacts that would arise from single-point or diagonal spanning choices [14,17].

Clusters were identified using a union-find algorithm following Hoshen-Kopelman logic [17]. Only occupied sites contribute to cluster identification, and nearest-neighbor connectivity was used throughout.

### 3.3. Quenched-disorder averaging

To separate randomness in lattice geometry from occupation noise, a two-level averaging protocol was used. For each parameter set, $N_G$ independent RhFD geometries were generated. For each geometry $G_a$, $N_O$ independent site-occupation realizations were performed. The ensemble-averaged spanning probability was estimated as

$$P_{perc}(p, \rho_h) = (1/N_G) sum_a (1/N_O) sum_b I_{\text{span}}(G_a, O_{a,b}).$$

Here $I_{\text{span}} = 1$ if spanning occurs and 0 otherwise. Deterministic baseline lattices were treated as a single geometry with repeated occupation trials.

### 3.4. Threshold and finite-cluster metrics

The percolation threshold $p_c(\rho_h)$ was estimated by linear interpolation at $P_{\text{perc}} = 0.5$. The finite-cluster-size response, excluding spanning clusters, was computed as

$$S_{\text{finite}} = (sum_s s^2 n_s)/(sum_s s n_s),$$

where $n_s$ is the number of non-spanning clusters of size $s$. The peak of $S_{\text{finite}}$ provides an independent indicator of the critical region.

### 3.5. Activation susceptibility

To quantify the sensitivity of the threshold to out-of-plane growth, the activation susceptibility was defined as

$$\chi_{\text{act}}(\rho_h) = -dp_c/d\rho_h.$$

Numerical derivatives were computed using finite differences on the discrete $p_c(\rho_h)$ values. This measure highlights endpoint-compressed activation of effective three-dimensional connectivity.

### 3.6. Computational implementation

All simulations were implemented in Python using NumPy-based lattice generation and union-find cluster identification. Parallelization was achieved through multiprocessing, with the number of CPU cores specified in an external parameter file. Visualization scripts produced three-dimensional lattice renderings and layer-occupation profiles for Figures 1 and 5.

## 4. Discussion

This study introduces random-h fractional dimension (RhFD) as a physically explicit lattice construction for non-integer-dimensional statistical mechanics. Unlike analytic continuation or abstract fractal dimension, RhFD generates lattices with well-defined sites, nearest-neighbor connectivity, and heterogeneous out-of-plane growth. The integer-dimensional endpoints reproduce the classical two-dimensional square-lattice and three-dimensional cubic-lattice percolation behavior, confirming that the construction and percolation protocol are robust. The model therefore provides a concrete substrate on which classical statistical-mechanical processes can be simulated between integer-dimensional limits.

The central finding is endpoint-compressed activation of effective three-dimensional connectivity. Across the 2D-to-3D interval, the percolation threshold decreases gradually at low and intermediate $\rho_h$, but the largest change occurs near the three-dimensional endpoint. This behavior shows that RhFD is not a simple linear interpolation between a square lattice and a cubic lattice. The system can acquire additional out-of-plane mass over a broad range of $\rho_h$, yet the percolative benefit of this added structure remains delayed until the random lattice becomes sufficiently saturated in height. In other words, geometric thickening precedes effective three-dimensional connectivity.

A key observation supporting this interpretation is the non-monotonic behavior of the coordination descriptor $D_h$. In regular integer-dimensional lattices, average coordination and spatial dimension are closely related: a square lattice has an interior coordination number of 4, whereas a simple-cubic lattice has an interior coordination number of 6. It is therefore tempting to use <h>/2 as a direct dimension-like descriptor. However, the RhFD results show that this interpretation is only partially valid in random fractional-dimensional lattices. As $\rho_h$ increases from the two-dimensional limit, the mass dimension $D_m$ increases because additional out-of-plane sites are added, but $D_h$ initially decreases below the two-dimensional baseline. This decrease does not mean that the system becomes globally "less than two-dimensional." Rather, early out-of-plane growth produces many terminal or weakly connected protrusions. These sites add mass but contribute relatively few nearest-neighbor bonds, thereby lowering the average coordination.

This behavior has two complementary implications. If $D_h$ is regarded as a dimension-like quantity, then its delayed and non-monotonic response demonstrates that the 2D-to-3D crossover is intrinsically non-linear. The system does not pass from two to three dimensions through a uniform increase in local coordination. Instead, the extra dimension becomes locally effective only after random growth has produced a sufficiently dense out-of-plane backbone. Conversely, if $D_h$ is regarded more cautiously as a local coordination descriptor rather than a true dimension, its initial decrease shows why average coordination alone cannot serve as a complete measure of effective dimensionality. It is strongly affected by protrusions, dangling ends, surface sites, and finite-height disorder, all of which are abundant in the intermediate RhFD regime.

The rapid increase of $D_h$ near $\rho_h = 1$ is therefore particularly informative. It marks the transition from a protrusion-dominated regime to a backbone-dominated regime. In the former, out-of-plane sites exist but remain locally inefficient for connectivity; in the latter, adjacent columns and layers become sufficiently populated to create redundant three-dimensional paths. This delayed rise in $D_h$ occurs together with the endpoint-compressed decrease of $p_c$, suggesting that local coordination and global percolative connectivity become activated in the same high-$\rho_h$ region. Thus, the RhFD crossover separates three notions that coincide more closely in regular lattices: mass growth, local coordination, and critical connectivity.

This separation is central to the physical meaning of RhFD. A single scalar "dimension" cannot fully characterize a random non-integer-dimensional lattice. The mass dimension $D_m$ measures how many sites exist; the coordination descriptor $D_h$ measures how efficiently these sites are locally connected; and the percolation threshold $p_c$ measures whether local connectivity has organized into a system-spanning backbone. Their mismatch is not a defect of the model, but one of its main findings. It reveals that effective dimensionality in a random lattice is not simply a state variable determined by mass or average coordination alone. Instead, it depends on how added sites are organized into connected pathways.

The results also suggest that fractional dimensionality in physical lattices may be construction-path dependent. In the present bottom-up RhFD construction, out-of-plane layers are generated recursively, so higher layers can exist only if the layers below them survive. This produces layer occupation profiles that decay approximately as $\Theta_z \sim \rho_h^z$, causing high layers to remain sparse until $\rho_h$ is close to one. A different construction, such as top-down depletion from a full three-dimensional lattice, may produce a different layer profile and therefore a different path through the same nominal dimensional interval. This possibility reinforces the idea that non-integer-dimensional lattices should not be characterized solely by endpoint dimensions or average mass scaling; the generative rule and internal connectivity architecture may also determine their critical behavior.

Beyond site percolation, RhFD provides a framework for studying nonlinear processes and complex systems on fractional-dimensional lattices, including Ising or Potts models, random walks, diffusion, and elastic networks. The present percolation study establishes the minimal connectivity problem. Future statistical-mechanical models can test whether the same separation between geometric mass, local coordination, and critical response also appears in thermal phase transitions, transport processes, and dynamical activation. In this sense, RhFD is not only a way to construct non-integer-dimensional lattices, but also a tool for examining how different physical meanings of dimension emerge, decouple, and become activated in complex systems.

## 5. Conclusion

We developed random-h fractional dimension (RhFD) as an explicit lattice construction for statistical mechanics in non-integer dimensions. Quenched site-percolation simulations demonstrate that the two-dimensional and three-dimensional integer endpoints are recovered, while intermediate lattices exhibit a robust 2D-to-3D crossover.

The percolation threshold decreases non-uniformly, producing endpoint-compressed activation: geometric mass increases first, but effective three-dimensional connectivity emerges predominantly near the full-growth limit. Mass dimension, coordination, and percolative connectivity evolve distinctly, revealing that dimensionality in random lattices is multi-faceted. RhFD thus establishes a computational substrate for exploring fractional-dimensional complex systems and nonlinear connectivity phenomena.

## Data and code availability

The numerical data and analysis scripts generated in this study are available from the corresponding author upon reasonable request. The RhFD lattice generation, quenched site-percolation simulation, threshold extraction, figure-generation scripts, and visualization scripts were implemented in Python.

## Declaration of competing interest

The authors declare no competing interests.

## Acknowledgements

The work in financially supported by the YRI-FD Industrial Project (YRI-IP-25-01).

## Supplementary material

Supplementary Movie 1. Random-h fractional-dimensional lattice growth from sparse protrusions to a three-dimensional slab. The animation visualizes the RhFD construction by increasing $\rho_h$ from 0 to 1 using a fixed random field. The color encodes height along the $z$ direction.